\documentclass[final,5p,times,twocolumn]{elsarticle}
\usepackage{color, soul}
\usepackage{hyperref}
\UseRawInputEncoding
\journal{Journal of \LaTeX\ Templates}

\begin{document}
	
	\begin{frontmatter}
		
		\title{Complex magnetoelectric effect in PFN-PT/CoFe$_{2-x}$Zn$_x$O$_4$  bulk  particulate composites }% Force line breaks with \\

		\author[]{Mohammad Torabi-Shahbaz\corref{cor1}}%
		\cortext[cor1]{Corresponding author}
		\ead{torabishahbazmohammad@gmail.com}
		\author[]{Hossein Ahmadvand}
		\author[]{Hadi Papi}
		\author[]{Saeideh Mokhtari}
		\author[]{Parviz Kameli}
		\address{Department of Physics, Isfahan University of Technology, Isfahan 84156-83111, Iran}

		\begin{abstract}
			The structural, dielectric, magnetic, and magnetoelectric (ME) properties of particulate composites containing lead-iron niobate and lead titanate piezoelectric 0.94[PbFe$_{0.5}$Nb$_{0.5}$O$_3$]-0.06[PbTi$_{0.5}$O$_3$] (PFN-PT)  and Zn-substituted cobalt ferrite magnetostrictive CoFe$_{2-x}$Zn$_{x}$O$_4$ (CF$_{2-x}$Z$_{x}$O); 0.6(PFN-PT)/0.4(CF$_{2-x}$Z$_{x}$O), x=0, 0.025, 0.1, 0.2, 0.3 (with ratio of 60 Wt\% ferroelectric and 40 Wt\% ferrite); have been investigated. We investigated the ME voltage coefficient as a complex quantity for all composite samples using the dynamic piezomagnetic coefficient, $q^{ac}$=$\partial \lambda^{ac}/{\partial H}$. The results reveal that tuning the magnetostrictive phase has a strong effect on the real part of the ME voltage coefficient. Doping zinc into cobalt ferrite modified the magnetic properties of the magnetic phase, such as magnetic anisotropy and coercive field, and hence the ME properties. The highest ME coefficient value of 12.33 $\frac{mV}{cm. Oe}$ was obtained for x=0.1 at the magnetic field of 755 Oe. In addition, the magnetic field at which the maximum value of the ME coefficient was observed ($H_{peak}$) strongly depends on the value of Zn substitution. The results were interpreted using the magnetic field dependence of the CF$_{2-x}$Z$_{x}$O magnetostriction.

		\end{abstract}
		
		\begin{keyword}
			\texttt{Multiferroics, Magnetoelectric effect, Piezoelectrics, Magnetostriction, Cobalt ferrite}
		\end{keyword}
		
	\end{frontmatter}
	
	%\linenumbers

	\section{INTRODUCTION}\label{Int}
	Multiferroics are a class of materials that have at least two ferroic orders (ferroelectric, magnetic, and elastic). Magnetoelectric (ME) materials are a subset of multiferroics in which ferroelectric and magnetic orders are coupled. The ME materials are classified into single-phase and composite groups in which composites show higher ME coupling values than single-phase ones\cite{spaldin2019advances,HansSchmid1994,palneedi2016status,Spaldin391,hadi,Feibeg,Kimura,eerenstein2006w}. Since in this class of materials magnetization (polarization) can be controlled by an/a electric field (magnetic field), they can be used in both electrical and magnetic areas, and have a wide range of applications, including magnetic field sensors, spintronics, data storage, actuators, photovoltaic solar cells, etc\cite{bibes2008towards,vopson2015fundamentals,Maignan,olabi2008design}.
	
	The ME voltage in composite magnetoelectric materials is generated by polarization change in the piezoelectric phase as a result of dynamic mechanical deformation of the magnetostrictive phase induced by the applied ac magnetic field in the presence of the dc magnetic field. In addition, the ME coupling coefficient is affected by magnetostriction and piezoelectric behavior of magnetic and electric phases, respectively. As a result, significant magnetostriction and piezoelectric coefficients are essential for obtaining a high ME voltage coefficient.
	
	Many studies have been conducted on piezoelectric-ferrite composites like:
	(BiBa)(FeTiZn)O$_{3}$/CoFe$_{2}$O$_{4}$\cite{papi2021dielectric}, PFN-PT/(Co,Ni)Fe$_{2}$O$_{4}$\cite{Mokhtari}, PMN-PT/CoFe$_{2}$O$_4$\cite{zabotto2020magnetodielectric}  BaTiO$_3$/CoFe$_{2}$O$_4$\cite{verma2019magnetic,etier2016direct}, PMN-PT/NiFe$_{2}$O$_4$\cite{ahlawat2013evidence}, PZT/Ni$_{0.2}$Co$_{0.8}$Fe$_{2}$O$_4$\cite{bammannavar2009magnetic},
	PZT/CoFe$_{2}$O$_4$\cite{atif2015interplay}, PVDF/CoFe$_{2}$O$_4$\cite{gonccalves2015development}, and etc.
	Among them, Pb-based ferroelectrics, such as PZT and  PMN-PT, have been frequently utilized as the ferroelectric phase in ME composites due to their excellent piezoelectric characteristics. For the magnetic phase, high magnetostrictive materials such as cobalt ferrite are commonly used in these composites.
	
	In this work, as piezoelectric material, we used the solid solution of lead-iron niobate and lead titanate;0.94Pb(Fe$_{1/2}$Nb$_{1/2}$)O$_3$-0.06PbTiO$_3$(PFN-PT)\cite{sitalo2011dielectric}. There is a ferroelectric-paraelectric (F-P) phase transition at approximately 125°C for this compound with x=0.06 ( morphotropic phase boundary)\cite{SINGH2010}. It also has the highest dielectric constant and piezoelectric coefficient, d$_{33}$, among other concentrations\cite{sitalo2011dielectric}. Possessing a high d$_{33}$ makes it a suitable candidate for use in magnetoelectric composites as a piezoelectric phase(d$_{33}$=600 $\frac{pC}{N}$ for x=0.06)\cite{sitalo2011dielectric}. However, the ME effect in a particulate composite of PFN-PT with magnetostrictive materials has been rarely reported\cite{Mokhtari}. On the other hand, the magnetic properties of magnetostrictive materials can be controlled by doping them. In this research, we doped zinc elements into cobalt ferrite to tune its magnetostriction properties. This let us study the affect of manipulation
	of the magnetic phase on magnetoelectric coefficient.
	
	In essence, in this work we select PFN-PT as the ferroelectric phase and zinc-doped CoFe$_{2}$O$_{4}$ (CFZO) as the magnetic
	phase. The structural, dielectric, magnetic, and ME voltage coefficients (a complex quantity) are
	investigated for the composites with different contents of Zn in 0.6(PFN-PT)-0.4(CF$_{2-x}$Z$_{x}$O).

	\section{EXPERIMENTAL DETAILS}\label{Exp}
	The polycrystalline solid solution of 0.94Pb(Fe$_{1/2}$Nb$_{1/2}$)O$_3$-0.06PbTiO$_3$(PFN-PT) was synthesized using the solid-state reaction method.
	The high purity raw materials of PbO (with 4Wt$\%$ excess of PbO to compensate for lead volatility during the heat treatment), Fe$_2$O$_3$, Nb$_2$O$_5$, TiO$_2$, and one Wt$\%$ Li$_2$CO$_3$ (to make a better structure of the perovskite phase and reduce electrical conductivity) were blended and ground for 1.5 h by hand in an agate mortar. The obtained powder was calcined at 900 $^{\circ}$C for four hours, and then reground and annealed in the air in two stages, 1070 and 1100 $^{\circ}$C for 4 and 6 h, respectively.
	
	Co(NO$_3$)$_2$.6H$_2$O, Fe(NO$_3$)$_3$.9H$_2$O, Zn(NO$_3$)$_2$.6H$_2$O, C$_6$H$_8$O$_7$.H$_2$O, and deionized water were used to synthesize nanoparticles of  CoFe$_{2-x}$Zn$_x$O$_4$(x=0,0.25,0.1,0.2,0.3) using standard sol-gel method. Then the resultant powder was calcined at 800 $^{\circ}$C for four hours. 
	
	Finally, the Zn-doped cobalt ferrite and PFN-PT were mixed to prepare (60Wt\%) PFN-PT - (40 Wt\%) CoFe$_{2-x}$Zn$_x$O$_4$ (x=0,0.25,0.1,0.2,0.3) particulate composites. The resultant powders were ground and prepared as pellets pressed at a pressure of 300 MPa and sintered at 800 $^{\circ}$C for four hours.
	
	The X-ray diffraction (XRD) patterns of the samples were measured at room temperature by using an Empyrean diffractometer with Cu - K$\alpha$ radiation. The microstructure of samples was analyzed using Field Emission Scanning Electron Microscopy (FE-SEM).
	Magnetization measurements of the composites were carried out by using a vibrating sample magnetometer (VSM) at room temperature. The relative permittivity of pure PFN-PT and other composites were measured at various frequencies (10 - 650 kHz) as a function of temperature using an LCR meter (IM3570 Hioki). Polarization-electric field (P-E) loops of the samples at room temperature were performed by using a homemade setup based on the Sawyer-Tower method.
	After polling the samples in an electric field of 10 kV/cm for one hour, the transverse magnetoelectric voltage coefficient was measured at room temperature by using the lock-in amplifier technique.

	\section{Results and discussion}
\begin{figure}[]
		\begin{center}
			\includegraphics[width=10cm,angle=0]{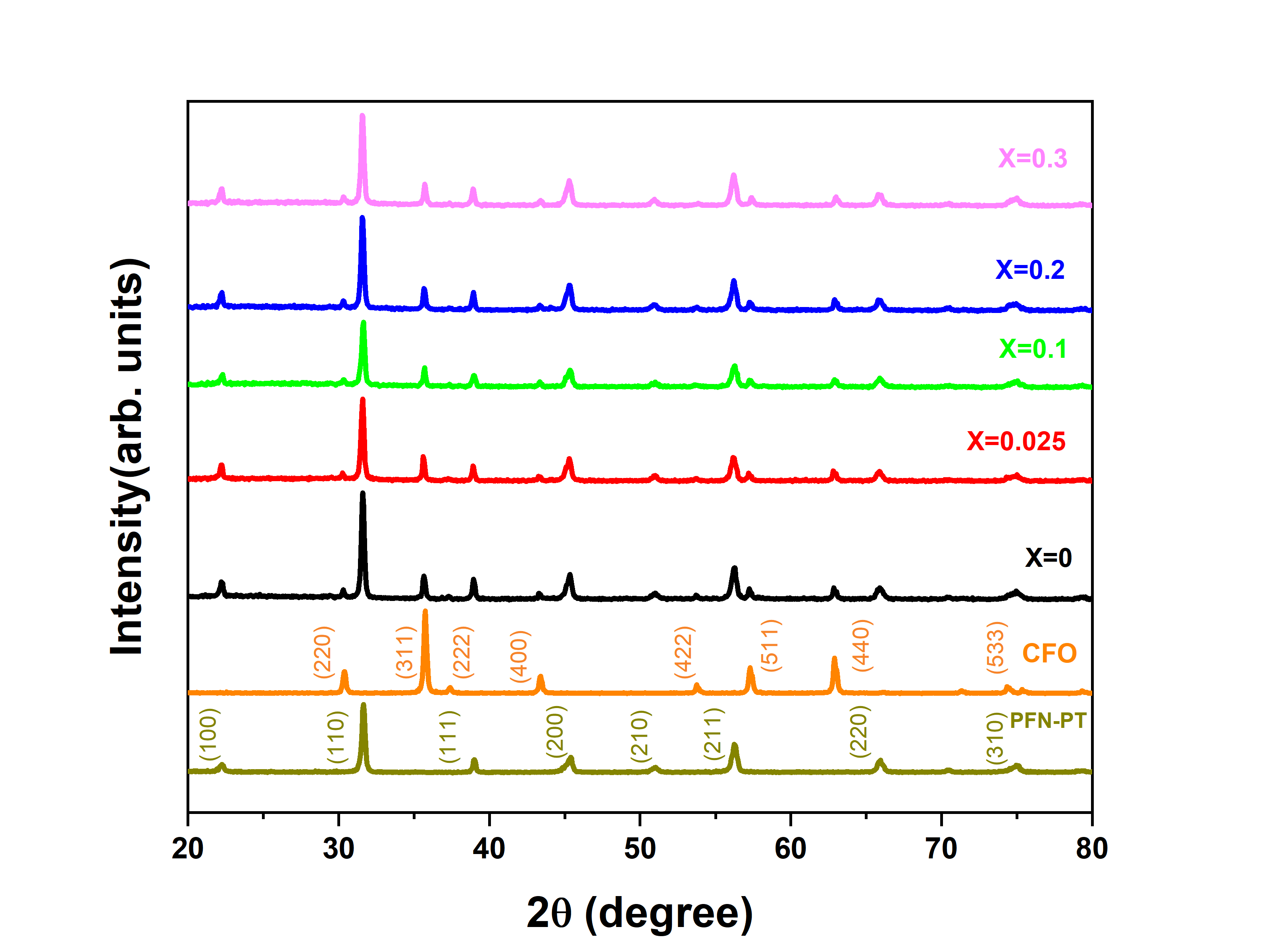}
			\caption{XRD patterns of (PFN-PT)-(CF$_{2-x}$Z$_{x}$O) composites (x=0, x=0.025, x=0.1, x=0.2, x=0.3) as well as pure PFN-PT and CFO.  
				\label{jadidxrdall}}
		\end{center}
	\end{figure}

    The XRD patterns of all composites, as well as polycrystalline PFN-PT solid solution and cobalt ferrite (CFO) samples (for comparison), are shown in figure. \ref{jadidxrdall}. The XRD pattern of the composite shows the structures of perovskite and spinel associated with PFN-PT and CFO, respectively. No impurity phase is observed in the XRD pattern of sintered composites which means that no significant chemical reactions occurred between the grains of ferrite and ferroelectric during the sintering.

\begin{figure}[]
		\begin{center}
			\includegraphics[width=9cm,angle=0]{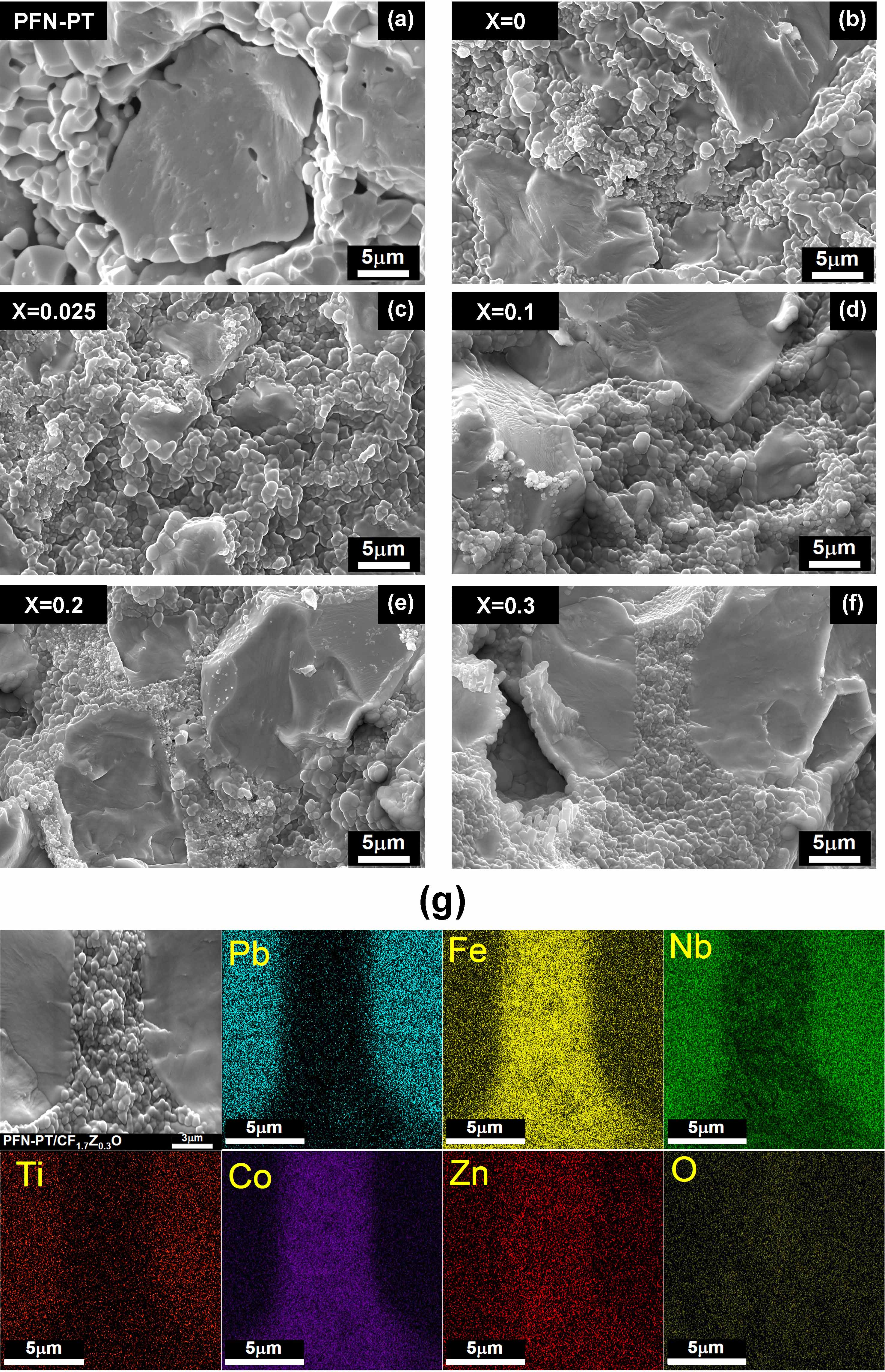}
			\caption{(a-f) FE-SEM images of the sintered pellets of the pure PFN-PT as well as  (PFN-PT)-(CF$_{2-x}$Z$_{x}$O) composites samples (with x=0, x=0.025, x=0.1, x=0.2, x=0.3). (g) EDS mapping images for the sample PFN-PT- CoFe$_{1.7}$Zn$_{0.3}$O$_4$ .
				\label{jadidfesem}}
		\end{center}
	\end{figure}

	Figure \ref{jadidfesem} shows the FE-SEM images of the sintered pellets of the pure PFN-PT (a), as well as all composite samples (b-f). It can be seen from the images that the samples consist of polygonal micrometer grains (related to the PFN-PT) and small CFO particles that are distributed in the shape of agglomerates inside the ferroelectric matrix. FE-SEM images exhibit that the ferrite particles are uniformly spread within the ferroelectric phase, which is suitable for achieving high magnetoelectric coupling.
	Figure \ref{jadidfesem} (g) shows the EDS mapping images of the sintered pellet of the sample PFN-PT/CoFe$_{1.7}$Zn$_{0.3}$O$_4$. The elements Pb, Nb, and Ti, are more prominent in areas of the main image where ferroelectric grains are present. On the other hand,  Co, Fe, and Zn elements are more distinguished in regions of the main image where there are tiny grains of ferrite.
	
\begin{figure*}[]
		\begin{center}
			\includegraphics[width=16.75cm,angle=0]{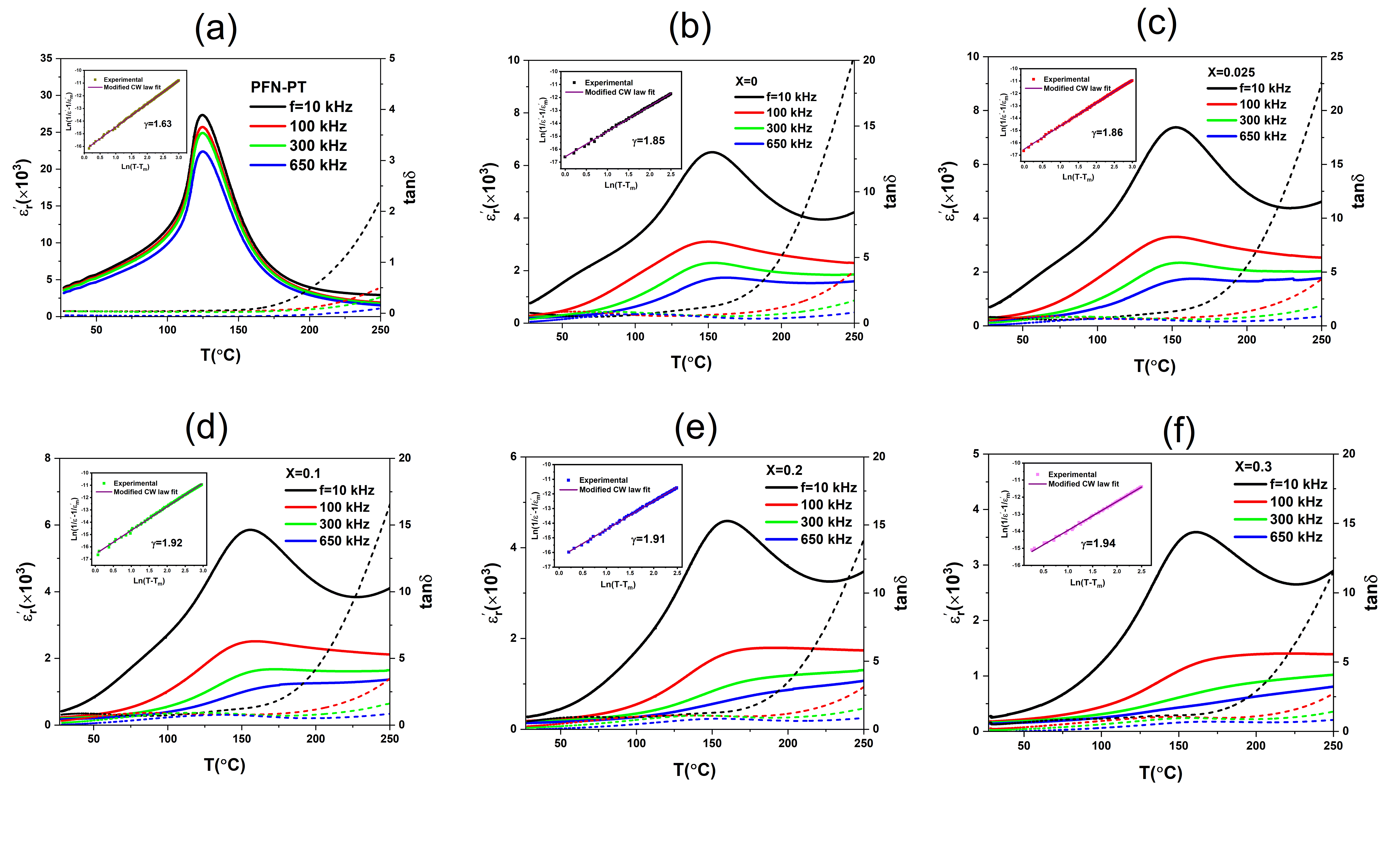}
			\caption{Temperature dependence of relative permittivity ($\epsilon^{\prime}$) (solid lines) and the corresponding dielectric loss tangent (tan$ \delta $) (dashed lines) in frequency range of 10 to 650 kHz for (PFN-PT)-(CF$_{2-x}$Z$_{x}$O) composites samples with (a) x=0, (b) x=0.025, (c) x=0.1, (d)
				x=0.2, (e) x=0.3, as well as pure PFN-PT (f).Insets show fits of data above transition temperature to modified Curie-Weiss law.   
				\label{allepsilon}}
		\end{center}
	\end{figure*}

	Figure \ref{allepsilon}  illustrates the temperature dependence of relative permittivity ($\epsilon_r$) and dielectric loss tangent (tan$ \delta $) at four frequencies (10, 100, 300, 650 kHz) in the temperature range of 25 - 250 $^\circ$C for pure PFN-PT (a) and composites samples (b-f). The dielectric constant of samples increases with rising temperature and reaches its maximum at $ T_{m} $ before decreasing with further temperature increases. As clearly seen from figure. \ref{allepsilon} (a), the dielectric constant of pure PFN-PT has a peak at 124.5 $^{\circ}$C. At this temperature, the dielectric constant has its maximum value and a broad peak, related to a phase transition from high-temperature paraelectric (cubic structure) to the low-temperature ferroelectric phase (tetragonal structure). Figures \ref{allepsilon} (b-f) show that by adding the magnetic phase, the dielectric constant ($\epsilon_r$) decreases and the loss tangent (tan $ \delta $) increases. This is owing to the decreased resistance of the ferrite magnetic phase as opposed to PFN-PT. The peak of the paraelectric to ferroelectric transition is also found to be moved to higher temperatures. The interdiffusion of ions between the magnetic and ferroelectric phases during the sintering process is responsible for the shift in transition temperature \cite{lin2009dielectric,schileo2018multiferroic}. As shown in figure \ref{allepsilon} (a), $ T_{m} $ for PFN-PT is frequency independent, but it moves toward higher temperatures as frequency increases for composite samples (see figure \ref{allepsilon} (b-f)).
	
	At temperatures above the transition temperature, the real dielectric constant follows the modified Curie-Weiss equation.
	\begin{equation}
		\frac{1}{\epsilon^{\prime}}-\frac{1}{\epsilon_m^{\prime}}=\frac{(T-T_{m})^{\gamma}}{C_{1}}
		\label{gammaeq}
	\end{equation}
	
	where $\epsilon_m^{\prime}$ is the highest dielectric permittivity at the transition temperature, $C_{1}$ is the modified Curie-Weiss constant, and $\gamma$  is the slope of log(1/$\epsilon^{\prime}$-1/$\epsilon_m^{\prime}$) vs. log ($T-T_{m}$). The value of $\gamma$ indicates the ferroelectric's degree of relaxation. Its value is 1 for normal ferroelectrics, but for relaxor ferroelectrics, $\gamma$ ranges between 1 and 2,  whereas ($\gamma$= 2) indicates the entirely diffuse phase transition (1$\leq$ $\gamma$ $\leq$ 2) \cite{xi2015effect,jan2018study,usman2013magnetic}. The graphs of ln(1/$\epsilon^{\prime}$-1/$\epsilon_m^{\prime}$) vs. ln ($T-T_{m}$) at 10 kHz are shown in the insets of figure \ref{allepsilon} (b–e). When the experimental data are fitted to Equation \ref{gammaeq}, the values $\gamma$=1.63, 1.85, 1.86, 1.92,1.91 and 1.94 were obtained for PFN-PT, x=0, 0.025, 0.1, 0.2, and 0.3 respectively.These results, when compared to pure PFN-PT with a smaller gamma, demonstrate typical relaxor behavior in the composite samples. Because of composition fluctuations and structural disorder in the cationic configuration on the sites of the crystal structure, this behavior is linked to the emergence of polar nano regions (PNRs), which are responsible for this behavior\cite{jan2018study}.

\begin{figure}[]
		\begin{center}
			\includegraphics[width=9cm,angle=0]{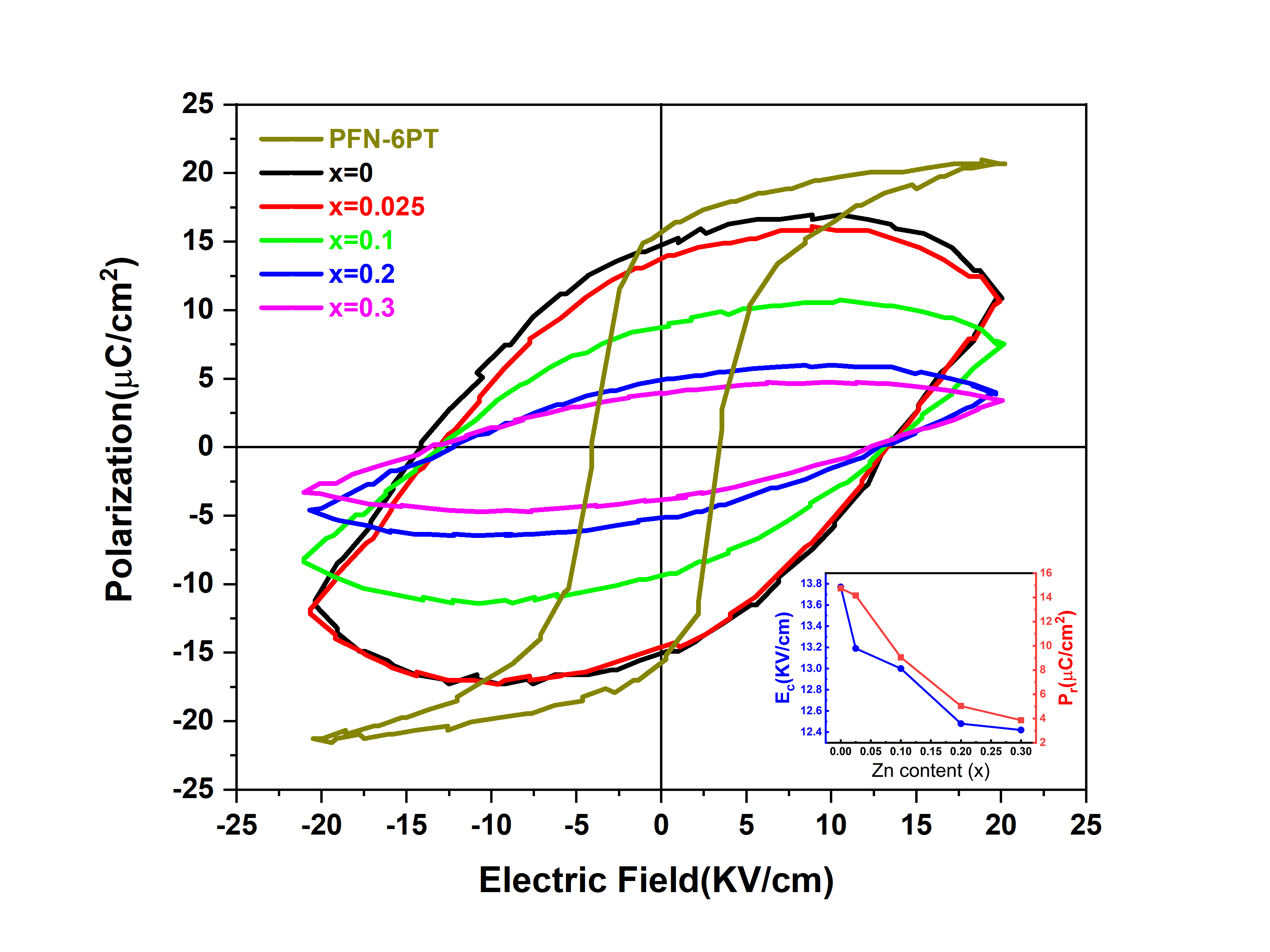}
			\caption{Room temperature polarization-electric field (P-E) loops of (PFN-PT)-(CF$_{2-x}$Z$_{x}$O) composites samples, as well as pure PFN-PT, measured at a frequency of 50 Hz. Inset shows the coercive field (Ec) and remnant polarization (Pr)
				as a function of Zn content. 
				\label{allp-eloop}}
		\end{center}
	\end{figure}

	Figure \ref{allp-eloop} exhibits the ferroelectric hysteresis loops of all samples up to the electric field of 20 kV/cm at a frequency of 50 Hz. Pure PFN-PT shows a coercive field (E$_c$) of 3.75 kV/cm and remanent polarization (P$_r$)  of 16 $\mu \textrm{C/m$^2$}$. E$_c$ and P$_r$ for composite samples are plotted in the inset of figure \ref{allp-eloop} as a function of Zn content. According to the graph, introducing the non-ferroelectric phase of ferrite to the ferroelectric phase PFN-PT reduces remanent polarization while increasing the coercive field. It is also noted that as the electric field increases due to the increased conductivity and leakage current of the ferrite magnetic phase, the composite samples are not as well-saturated as the pure sample and the loops form becomes rounder. The E$_c$ value is nearly identical for all composite samples. A high E$_c$ indicates that a larger electric field is required to rotate the orientation of the electric domains and make the sample's total polarization zero. It is primarily due to porosity within the material and the space charges that generate electric domain wall pinning at ferrite and ferroelectric grain boundaries\cite{uchino2018ferroelectric,gupta2011improved,tyagi2015large}.
	
	Also, the quantity of remanent polarization in the composite samples varied according to the amount of zinc impurity in the magnetic phase. As the amount of zinc in the magnetic phase increases, the remanent polarization of the composite drops. According to the dielectric properties data, this decrease in polarization can be attributed to the higher conductivity of samples with  a higher zinc content. As conductivity increases due to electric leakage current, a lower quantity of surface charge accumulates on the electrodes, reducing electric polarization.
	
\begin{figure}[]
		\begin{center}
			\includegraphics[width=9cm,angle=0]{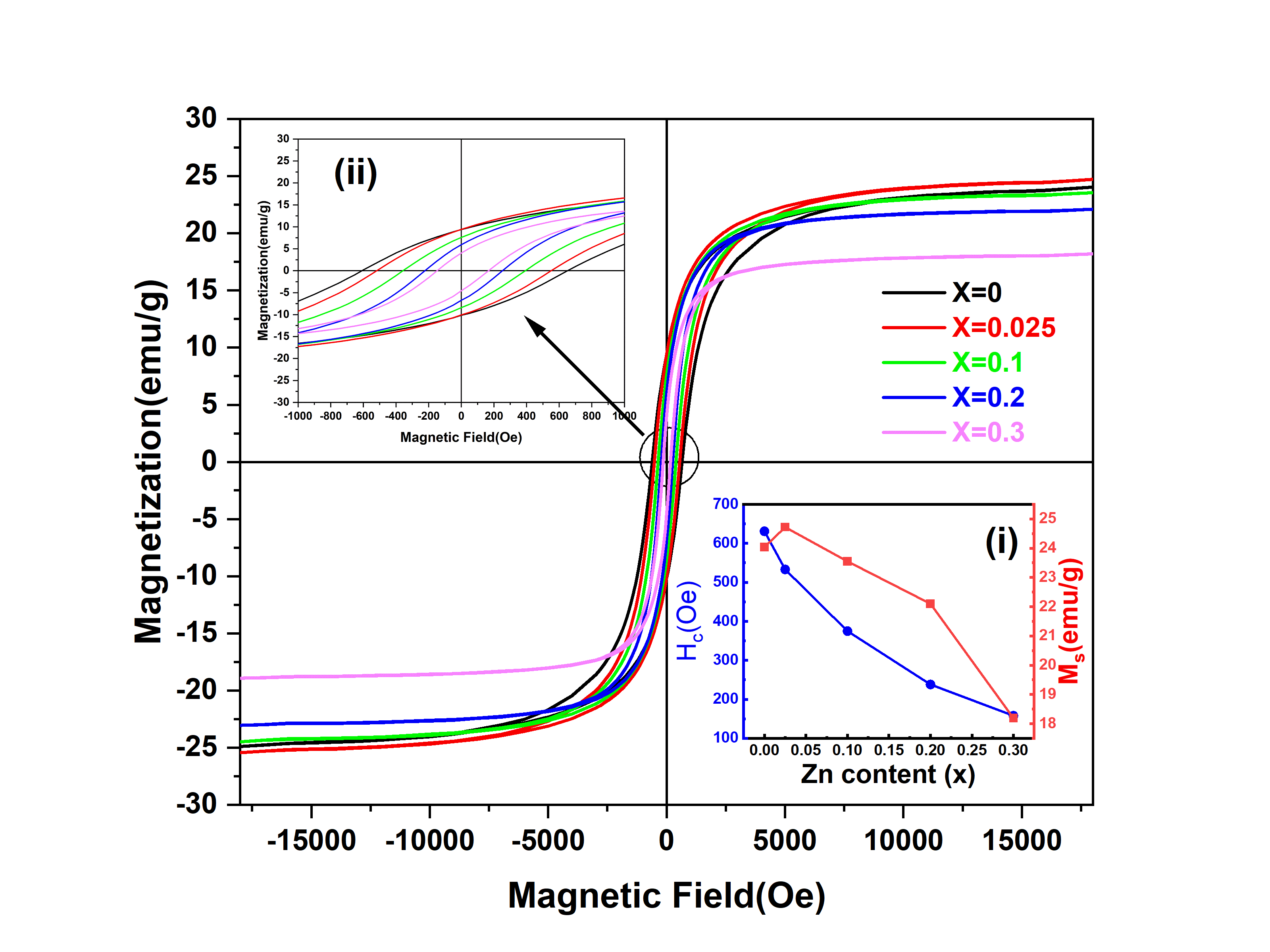}
			\caption{Room temperature magnetization loops of (PFN-PT)-(CF$_{2-x}$Z$_{x}$O) composites. Inset (i) shows the coercive field (H$_c$) and saturation magnetization (M$_s$) as a function of Zn content. Inset (ii) shows a zoomed view of the m-h loop in the vicinity of the center.
				\label{malekmhloop}}
		\end{center}
	\end{figure}

	The magnetic hysteresis curves for all composite samples are shown in figure \ref{malekmhloop}. Because the PFN-PT is paramagnetic at ambient temperature, it contributes only a minor amount to the magnetic hysteresis curves, and the main contribution is related to zinc-doped cobalt ferrite. The saturation magnetization ($ M_{s} $) and coercive field ($ H_{c} $) value change with zinc impurity, as shown in the inset of figure \ref{malekmhloop}.
	
	For composite samples, the value of $ M_{s} $ grows and then drops. The value of $ M_{s} $ is determined by the distribution of zinc in the tetrahedral and octahedral sites. According to Neel's model, the net magnetization of spinel ferrites is equal to the difference in magnetizations in two sublattices $ A $ and $ B $, which represent the tetrahedral and octahedral sites, respectively. The $ Zn^{+2} $ ion preferentially occupies tetrahedral sites. Because the $ Zn^{+2} $ ion is non-magnetic and has no magnetic moment, by zn doping the magnetization value in sublattice $ A $ drops when compared to the magnetization value in sublattice $ B $, causing the net magnetization to increase. When the $ Zn $ content increases, some $ Zn^{+2} $ ions are placed in the octahedral site, causing magnetization to decrease further. The leading cause of the decrease in magnetization is the weakening of the superexchange interaction between the A and B sites, as well as the tilting of the moments\cite{singhal2010effect}.
	
	The $ H_{c} $ decreases with increasing zinc impurity in the iron site, as shown in the inset of figure \ref{malekmhloop}. The $ H_{c} $ in spinel ferrites is affected by parameters such as anisotropy constant, saturation magnetization, superexchange interactions, and lattice defects\cite{rao2013improved}. The  $ H_{c} $ is directly related to anisotropy. Anisotropy reduces as zinc impurity increases. As a result, the coercive field decreases. It is significant to mention that during the heating process in the furnace, atoms of ferroelectric and ferrimagnetic phases can penetrate each other, affecting the magnetic properties of ferrite.
	
	Magnetoelectric coupling in composites is primarily created by transmitted stress caused by magneto-mechanical interaction at the interface of the piezoelectric and magnetostrictive phases\cite{velev2011multi}. The magnetoelectric voltage coefficient is stated as a complex quantity\cite{Mokhtari}:
	
	\begin{equation}
		\alpha_{ME}^{\ast}=\alpha_{ME}e^{-i\varphi}=\alpha^{\prime}-i\alpha^{\prime\prime}
	\end{equation}
	
	where $\varphi$ is the phase shift between the driving current for the 
	ac magnetic field and the output ME voltage from the sample. The real and imaginary parts of the magnetoelectric coefficient are represented by $\alpha^{\prime}$ and $\alpha^{\prime\prime}$, respectively. This phase shift results in the formation of an imaginary portion, which represents the presence of energy loss during the measurement of $\alpha_{ME}^{\ast}$ for ME composites. A detailed explanation of ME measurement is described in the supplemental material of Ref\cite{Mokhtari}.
	
\begin{figure}[]
		\begin{center}
			\includegraphics[width=9cm,angle=0]{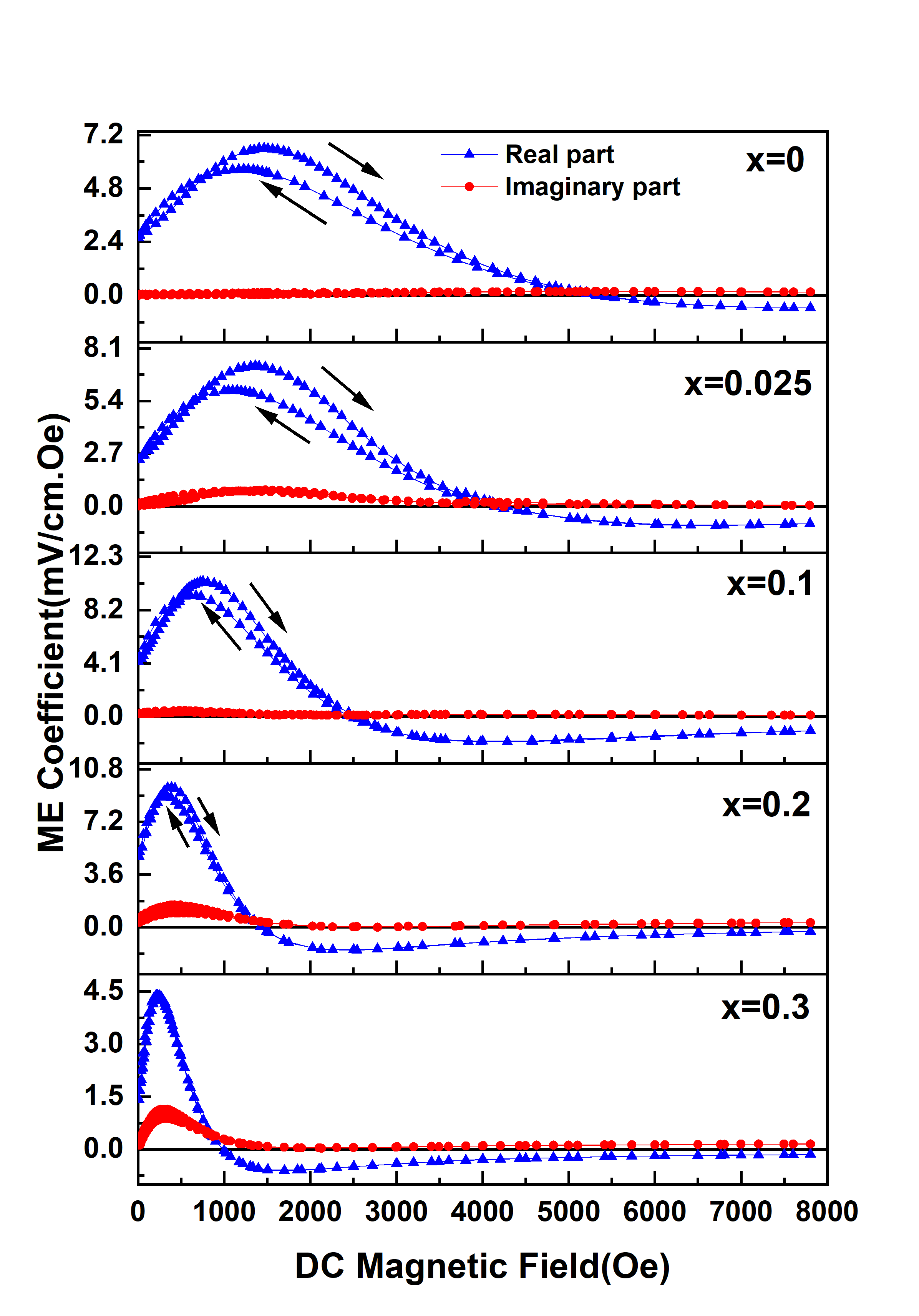}
			\caption{Real and imaginary part of transverse ME voltage coefficient ($\alpha^*_{31}$) as a function of DC magnetic field for (PFN-PT)-(CF$_{2-x}$Z$_{x}$O) composites (x=0, x=0.025, x=0.1, x=0.2, x=0.3), h$_{ac}$=0.77 Oe, and f=978 Hz.  
				\label{meallcfzo}}
		\end{center}
	\end{figure}

	To explore the influence of zinc substitution on magnetoelectric characteristics, after polling samples up to the $10 \frac{kV}{cm}$, the complex magnetoelectric coefficient was evaluated for composites. Figure \ref{meallcfzo} reveals the real and imaginary parts of ME  voltage coefficient for the composites as a function of DC magnetic field ($H_{DC}$) from 0 to 8 kOe in a 0.77 Oe AC magnetic field (f=978 Hz).
	
	In the absence of a direct magnetic field ($H_{DC}$), the real part of the ME coefficient has a non-zero value ($\alpha_{self-biased}$ ) in all samples, which is the result of the interaction between the alternating magnetic field ($H_{AC}$) and remanent magnetization of the ferrite phase.
	As seen in the figures, the magnetoelectric coefficient grows nearly linearly with increasing magnetic field ($H_{DC}$) for all samples until a specific magnetic field ($H_{peak}$) reaches its maximum value ($\alpha_{max}$). The magnetoelectric coefficient then decreases as the magnetic field increases until it reaches zero in a specific magnetic field. As the field strength increases, the value of the real magnetoelectric part changes sign and becomes negative. The values of $\alpha_{self-biased}$, $\alpha_{max}$, and $H_{peak}$ for all composite samples are presented in table \ref{metable}.

	\begin{table}
		\begin{center}
			\begin{tabular}{|c|c|c|c|}
				\hline
				{sample name} & $\alpha_{{self-biased}}(\frac{mV}{cm.Oe})$ &  $\alpha_{{max}}(\frac{mV}{cm.Oe})$& ${H}_{{peak}}(Oe)$ \\	\hline
				$ x=0 $& 2.57 & 7.16&1465\\ \hline
				$ x=0.025 $& 2.38 & 8.09&1360\\ \hline
				$ x=0.1 $& 4.20 & 12.33&755\\ \hline
				$ x=0.2 $& 4.79 & 11.2&388\\ \hline
				$ x=0.3 $& 1.41 & 4.96 &249\\ \hline	
			\end{tabular}\label{metable}
			\caption{The values of $\alpha_{self-biased}$, $\alpha_{max}$, and ${H}_{peak}$ for all composites.}
			
		\end{center}
	\end{table}

	 As shown in the figures, the values of $\alpha^{\prime}$ and ${H}_{peak}$ strongly depend on the zinc content in the cobalt ferrite magnetic phase. According to figure \ref{meallcfzo}, the value of $\alpha_{max}$ grows with increasing Zn content up to x = 0.1 and thereafter drops. Similar to coercivity, increasing Zn content causes ${H}_{peak}$ to shift toward lower fields. Sample x=0.1 has the optimal amount of Zn and the most robust magnetoelectric coupling.

	Another significant feature visible in the curves is hysteresis behavior between the increasing and decreasing branches of the $\alpha^{\prime}$. This hysteresis behavior reflects the magnetic hysteresis behavior. As shown in the pictures, as the Zn amount increases, the hysteresis curves become narrow, which is consistent with the magnetic hysteresis behavior of the samples.
	
	In general, the magnetoelectric voltage coefficient in composites depends on the magnetostriction ($\lambda$) and piezomagnetic coefficient $(\frac{\partial\lambda}{\partial M})$ of the magnetostrictive phase \cite{Agarwal2012,duong2007driving}. The ME effect is proven to be driven by dynamic piezomagnetic, $q^{ac}$, rather than quasi-static magnetostriction\cite{aubert2018dynamic}. $q^{ac}$ is described by:

	\begin{equation}
		q^{ac}=\frac{{\partial\lambda^{ac}}}{{\partial H_{ac}}}|H_{dc}
	\end{equation} 
	
	Where $q^{ac}$ is the dynamic magnetostriction induced by the vibration of magnetic moments by $H_{ac}$ while applying a constant DC magnetic field\cite{aubert2018dynamic}. Because of dynamic magnetostriction in the magnetostrictive phase, $\alpha_{self-biased}$ appears in the samples. The self-biased ME is due to the magnetic hysteretic response of the piezomagnetic phase\cite{ahlawat2017electric}. As a result, even without a direct magnetic field, all samples exhibit $\alpha_{self-biased}$.
	
	The variation in magnetostriction behavior among the composites in the presence of a magnetic field accounts for the variance in magnetoelectric behavior. Many investigations have been published on the magnetostriction curves of pure cobalt ferrite and cobalt ferrite doped with zinc\cite{somaiah2012magnetic,nlebedim2014dependence,anantharamaiah2017tuning}. The magnetostriction of zn-doped cobalt ferrite increases initially and decreases with increasing magnetic field. The magnetostriction of a zn-doped cobalt ferrite polycrystal sample is a combination of $\lambda_{100}<0 $ and $\lambda_{111}>0 $, where $\lambda_{100}$ and $\lambda_{111}$ correspond to magnetostriction in the easy and hard magnetic directions, respectively. The dominant contribution of $\lambda_{100}$ drives the increasing magnetostriction behavior in low fields. In contrast, the decreasing behavior of magnetostriction in high fields is due to the higher contribution of $\lambda_{111}$\cite{nlebedim2010effect,atif2013influence,anantharamaiah2016enhancing}. As the amount of the zinc element in the iron sites increases, the peak value of the magnetostriction curve decreases. Zinc substitution in cobalt ferrite causes part of magnetization vectors to change their directions from the easy axis (100) to the hard axis (111). Less magnetocrystalline anisotropy is a consequence of these changes in the direction of the magnetization vectors, which is in agreement with the results of the magnetic hysteresis curves. It turns out that the Zn-substituted samples magnetostriction peak (x=0.025,0.1,0.2,0.3) occurs at lower magnetic fields than the sample without Zn (x=0)\cite{somaiah2012magnetic,anantharamaiah2017tuning}.
	
	The value of $\alpha^{\prime}$, as previously mentioned, is proportional to the strain sensitivity $(\frac{\partial\lambda}{\partial M})$. Indeed, the larger the slope of the magnetostriction curve versus the magnetic field, the larger the coupling coefficient. In this investigation, the sample with the highest magnetoelectric coupling is $X=0.1$. In  prior investigations, the strain sensitivity of ferrite $CoFe_{1.9}Zn_{0.1}O $ confirmed that this impurity had the highest strain sensitivity\cite{somaiah2012magnetic,anantharamaiah2017tuning}.
	
	In order to gain further insight, we can interpret the magnetic field dependency of $\alpha^{\prime}$ by using the $\lambda (H)$ trend. As was already discussed, $\lambda^{dc} $ tends to grow as the magnetic field increases. In the case of high fields, the behavior of $\alpha^{\prime}$  is controlled by the type of the magnetic phase and the percentage of zinc impurity. When the DC magnetic field is strong enough to rotate the magnetic domains (${H}_{{peak}}$), the ac magnetic field can cause the maximum length change in the magnetic phase, yielding the highest value of $\alpha^{\prime}$. With further increasing the magnetic field, the value of $\alpha^{\prime}$ becomes zero $(\frac{\partial\lambda}{\partial H}=0)$ and then the sign of $\alpha^{\prime}$ changes from positive to negative, which is due to the negative sign of $(\frac{\partial\lambda}{\partial H})$ for zinc-doped cobalt ferrite \cite{somaiah2012magnetic,anantharamaiah2017tuning}. There have rarely been reports on negative ME voltage coefficients\cite{etier2013magnetoelectric,naveed2017effect,srinivasan2005dynamic}.
	
	Figure \ref{meallcfzo} also includes the samples' imaginary part of the ME coefficient ($\alpha^{\prime\prime}$). ME responses have rarely been studied in terms of their imaginary part\cite{Zhou2008,zhai2007large,guo2013frequency}. The imaginary part of the ME coefficient represents the energy dissipation in the system. As can be seen from the figure, the $\alpha^{\prime\prime}$ has a peak close to the ${H}_{{peak}}$. As mentioned earlier, $\alpha_{max}$ is obtained when the magnetic domains collectively rotate, producing the highest energy dissipation in the magnetic phase. It implies that the peak of $\alpha^{\prime\prime}$ should occur in the vicinity of  ${H}_{peak}$, where the peak of the real part occurs. This is true for all samples, as evidenced by the curves.
	
	\section{Conclusion}
	
	To sum up, the composite 0.94PbFe$_{0.5}$Nb$_{0.5}$O$_{3}$-0.06PbTiO$_{3}$ and CoFe$_{2-x}$Zn$_{x}$O$_{4}$  were prepared for  x=0, 0.025, 0.1, 0.2, 0.3 and a weight percentage of 60-40 (60 percent ferroelectric and 40 percent ferrite). The XRD characterization revealed the formation of perovskite and spinel structures for ferroelectric and ferrite phases, respectively. In addition, FE-SEM and EDS mapping images revealed a homogeneous and suitable distribution of the magnetic phase within the ferroelectric matrix. The ferroelectric to paraelectric phase transition occurs at 124.5°C in the dielectric constant curve for the pure PFN-PT. The dielectric constant is reduced by adding the ferrite phase. The electric hysteresis loop for the pure PFN-PT is well-saturated in the $ 20\frac{kV}{cm} $ field. The remanent polarization is reduced by including the non-ferroelectric ferrite phase.
	Furthermore, the value of the coercive field increases with the addition of ferrite, which is caused by the domain wall pinning. The magnetic hysteresis curves revealed that the coercive field value reduces as the zinc impurity increases. For each composite, the real and imaginary parts of the magnetoelectric coefficient were measured. All the composite samples possess $\alpha_{self-biased}$. The highest ME voltage coefficient value of 12.33 $\frac{mV}{cm.Oe}$ at DC magnetic field of 755 Oe was achieved for x = 0.1 sample. The measurements revealed that the magnetoelectric coefficient is highly dependent on the zinc content. In fact, with the substitution of the zinc in the iron sites, the strain value ($\lambda(H)$) and strain sensitivity $(\frac{\partial\lambda}{\partial H})$ of ferrite vary, and consequently, the value of the magnetoelectric coefficient changes based on these two parameters.

	\section{Acknowledgement}
	The Isfahan University of Technology supported this work
	(IUT).
	
	\section*{References}
	
	\bibliography{mybibfile3}

\end{document}